\documentclass[11pt]{article}
\usepackage{moriond,epsfig,amsmath,amssymb,amsfonts}

\def\Journal#1#2#3#4{{#1} {\bf #2}, #3 (#4)}

\def\PLB{{\em Phys. Lett.} B}
\def\PRL{\em Phys. Rev. Lett.}

\def\be{\begin{equation}}
\def\ee{\end{equation}}
\def\bea{\begin{eqnarray}}
\def\eea{\end{eqnarray}}

\begin{document}
\vspace*{4cm}
\title{HARD PROBES AT HEAVY-ION COLLIDER ENERGIES:\\ RESULTS FROM PHENIX.}

\author{ DAVID G. D'ENTERRIA for the PHENIX COLLABORATION }

\address{SUBATECH BP 20722, 44307 Nantes Cedex 3, Bretagne, France}

\maketitle\abstracts{
Hard processes in nucleus-nucleus interactions at relativistic energies 
are reviewed with emphasis on recent PHENIX results from the first run of the Relativistic 
Heavy-Ion Collider at BNL. The observed suppression of moderately high $p_T$ hadrons 
($p_T$ = 2 - 5 GeV/c) in $\sqrt{s_{NN}}$ = 130 GeV $Au+Au$ central collisions with respect
to the scaled $pp$ data, is discussed in terms of conventional nuclear and 
``quark-gluon-plasma'' effects. The meson and baryon composition at high $p_T$, 
as well as the implications for open charm of the measured single electron spectrum 
are also presented.
}

\section{Introduction.}

High-energy heavy-ion (HI) physics aims at the study of the QCD phase diagram at
energy densities where lattice calculations predict a phase transition from
hadronic matter to a (deconfined and chirally symmetric) plasma of quarks and
gluons (QGP). High-$p_T$ particles (jets, prompt $\gamma$) and heavy particles
($D$, $B$, $J/\Psi$) are produced in parton-parton scatterings with large momentum transfers 
(``hard processes'') during the earliest stages of a HI collision ($\tau\propto 1/p_T<$ 0.1 fm/c). 
Such hard probes are thus direct signatures of the partonic phases of the reaction 
and have attracted much interest because their yields can be quantitatively compared
(after scaling with the number of binary inelastic collisions $N_{coll}$)
with: (i) perturbative QCD predictions, and (ii) the available set of baseline ``vacuum'' 
($pp,p\bar{p}$) and ``cold-medium'' ($pA$) data. In both cases, any departure 
(suppression or excess) from the ``expected'' results provides precious information on 
the strongly interacting hot and dense medium created during the reaction.

\section{Run-1 PHENIX experimental setup.}

During the year 2000, $\sim$5 million ``minimum bias'' $Au+Au$ events ($\sim$1 $\mu$b of
integrated luminosity) at a center-of-mass energy of 130 GeV were collected by the
PHENIX experiment at Brookhaven National Laboratory RHIC collider.
The PHENIX detector is specifically designed to measure penetrating probes, such as direct photon
radiation, heavy flavors, and jet production, by combining good mass and PID resolution,
high rate capability and small granularity, though within a limited angular coverage.
PHENIX is composed of 11 detector subsystems divided into: (i) 2 central arm spectrometers for
electron, photon and hadron measurement at mid-rapidity ($|\eta|<0.38$, $\Delta\phi=\pi/2$); 
(ii) 2 forward-backward ($|\eta|=1.15 - 2.25$, $\Delta\phi=2\pi$) spectrometers for 
muon detection; and (iii) 3 global (inner) detectors for centrality and trigger selection.
During this Run-1, neutral pions were reconstructed through invariant mass analysis of $\gamma$ pairs 
detected in the 3 active sectors of the electromagnetic calorimeter EMCal: two sectors of lead
scintillator (PbSc) on the west arm and one sector of lead glass (PbGl) on the east side.
In the east arm, trajectories and momenta of charged hadrons were measured by a drift chamber
(DC) and two MWPC's with pad readout (PC1 and PC3). Hadron identification was achieved with 
the time-of-flight detector (TOF) within a third of the azimuthal acceptance of the east
arm. Electron detection was done combining the ring-imaging-cerenkov (RHIC) and the EMCal
energy and momentum measurements.


\section{High-$p_T$ hadron yields. Conventional nuclear effects vs QGP signals.}

\begin{figure}
\psfig{figure=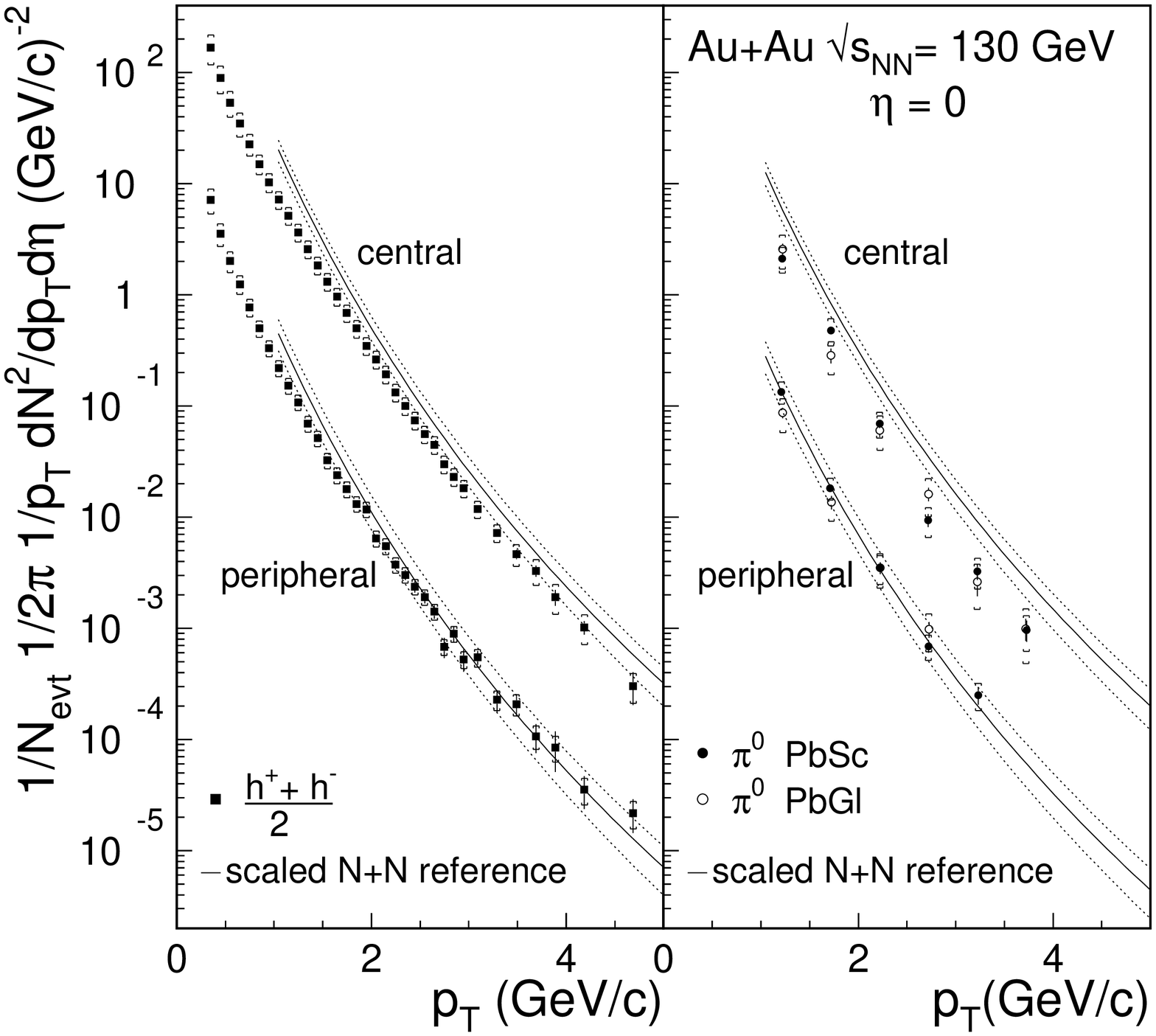,height=3.2in}
\psfig{figure=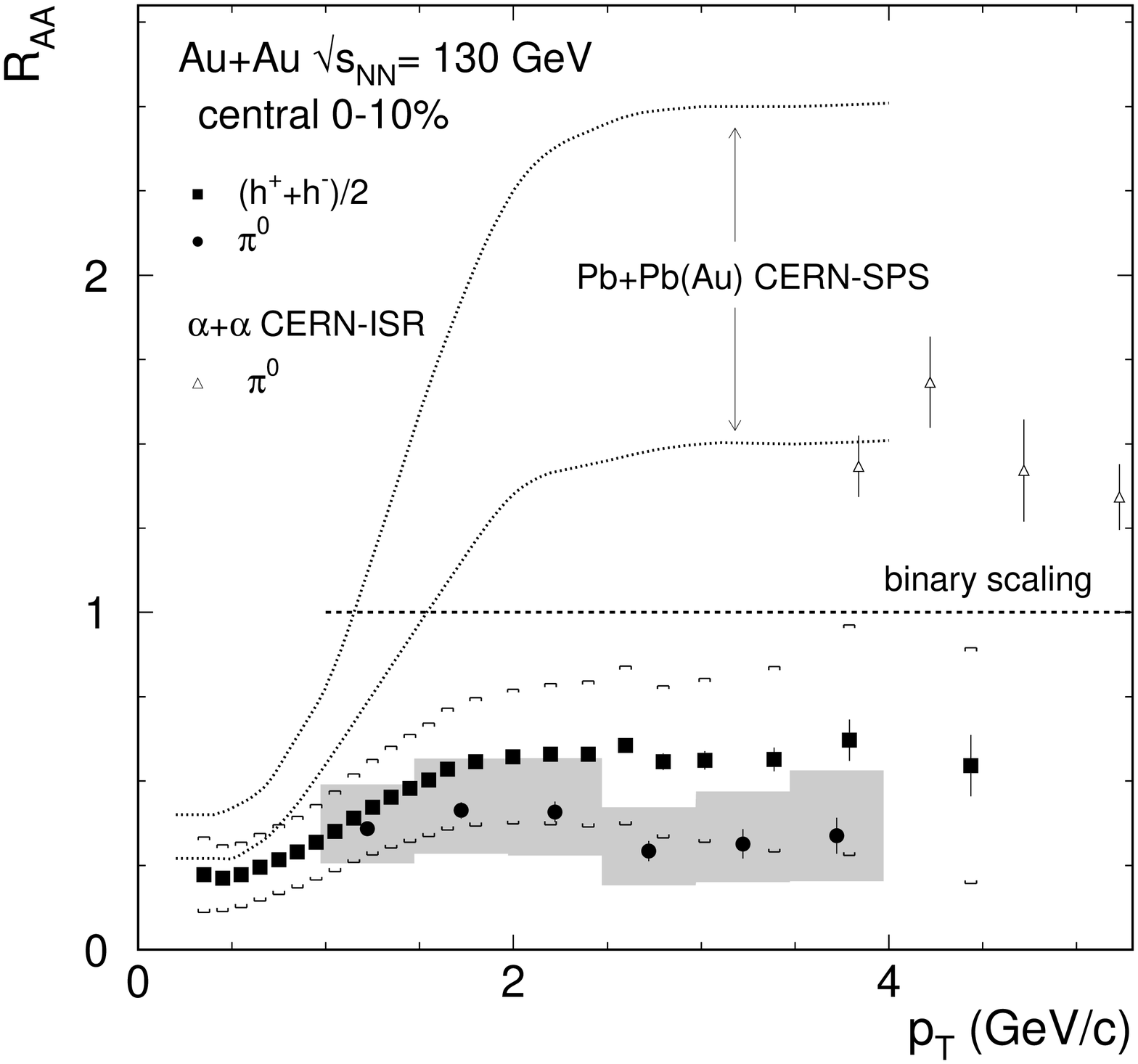,height=3.2in}
\caption{Left: Transverse momentum distribution of charged hadrons and $\pi^0$
measured by PHENIX \protect\cite{enterria:ppg003} in 10\% most central and
in 60\%-80\% peripheral $Au+Au$ collisions. The solid lines are the binary-scaled
extrapolation of $pp(\bar{p})$ data. Right: Nuclear modification factor, 
Eq. (\protect\ref{eq:nucl_modif_factor}), for central collisions compared 
to data at lower energies.
\label{fig:1}}
\end{figure}

Due to the large soft particle background produced in HI collisions, the standard jet algorithms
in $pp$ collisions fail to identify jets below $p_T\sim$ 40 GeV/c in $Au+Au$ at 
$\sqrt{s_{NN}}\sim$ 200 GeV. Yet, ``jet physics'' in HI collisions can still be studied 
via high transverse momentum hadrons ($p_T>$2 GeV/c) under the assumption that most of
these particles result from the fragmentation ({\it leading particle}) of 
the partons produced in the hard scattering process. Fig. \ref{fig:1} (left) shows 
the yield of charged hadrons and $\pi^0$ mesons ($p_T>$1.0 GeV/c) emitted at mid-rapidity in central 
and peripheral collisions compared to the scaled nucleon-nucleon ($NN$) reference. 
Whereas in peripheral collisions the yield is well reproduced by the binary-scaled 
extrapolation of $pp(\bar{p})$ (ISR, UA1, CDF) yields, in central collisions the yields 
are suppressed by a factor 2 to 3 \cite{enterria:ppg003}. This suppression is more visible 
in the right side of Fig. \ref{fig:1} where the ``nuclear modification factor''

\begin{equation}
R_{AA}(p_T)\,=\,\frac{(dN/dp_T)_{AA}}{\langle N_{coll}\rangle (dN/dp_T)_{NN}},
\label{eq:nucl_modif_factor}
\end{equation}

is plotted. The value of $R_{AA}(p_T)$ quantifies the deviation 
from the nucleon-nucleon extrapolation, i.e. from the absence of nuclear-medium effects,
in terms of suppression or enhancement ($R_{AA}$ smaller or larger 
than unity, respectively). For $p_T<$ 2 GeV/c, $R_{AA}$ is expected to be below
unity since the bulk of particle production is due to soft processes which
scale with the number of participant nucleons ($N_{part}$) rather 
than with $N_{coll}$. Above $p_T$ = 2 GeV/c, however, hadron production
in $pA$ and $AA$ collisions at lower energies \cite{enterria:kope02,enterria:WA98} 
is found to be enhanced compared to the binary scaling expectation. 
This increased high-$p_T$ production in the nuclear medium 
(``Cronin effect'') is due to initial-state multiple (soft \& semi-hard) parton 
interactions before the hard scattering, resulting in a $k_T$ broadening additional 
to the intrinsic nucleon $k_T$ observed in $pp$ collisions.
Notwithstanding, PHENIX observes exactly the opposite trend: instead of an 
enhancement, a suppression of the particle yield is observed above 2 GeV/c.

Apart from the Cronin enhancement, the other initial-state effect that is known to 
modify the $p_T$ distributions is nuclear shadowing. The region of parton fractional 
momenta $x_T=2p_T/\sqrt{s}$ relevant for moderately high-$p_T$ production (2-5 GeV/c) at RHIC is 
$x_T\sim$ 0.04 - 0.1, where gluons are known to dominate the parton distribution functions 
(PDFs) in the nucleon. In the nuclei, it has been experimentally observed that the 
PDFs are depleted compared to the nucleon ones for $x<$0.1. 
Such a shadowing is usually explained in terms of parton recombinations which effectively
reduce the parton density at low $x$ in the nuclei. The most up-to-date parametrizations 
of nuclear PDFs \cite{enterria:EKS} predict, however, a very small (if any) reduction of 
the gluon density in the range of $x$ and $Q^2$ values compatible with the high-$p_T$ PHENIX
results. This fact rules out shadowing as a possible explanation of the
observed hadron deficit\footnote{Note, however, that the ($x,Q^2$) scale dependence of 
nuclear PDFs (specially for gluons) is less known than for nucleons due to the 
limited range of $x, Q^2, A$ covered in the available DIS $lA$ experiments. Therefore,
$p,d+A$ collisions at the same $\sqrt{s}$ as the $Au+Au$ ones are needed at RHIC 
in order to 
assess the role of gluon shadowing.}.\\

The most interesting explanation of the high-$p_T$ suppression relies not on an initial-state
nuclear effect but on a final-state {\it medium} one: ``jet quenching''. It has been 
since long predicted \cite{enterria:Gyu90} that a hot and dense colored medium like a QGP would induce
multiple gluon radiation off the fast partons produced in a hard scattering.
The resulting energy loss of the parton, $dE/dx$ emitted outside the jet cone 
before its hadronization, is of the order of a few GeV/fm and would lead to a
depletion of the leading-particle inclusive spectra at high $p_T$. Leading-order
pQCD calculations \cite{enterria:XNW,enterria:Fai} including $k_T$ Cronin enhancement, nuclear PDFs, and modified 
fragmentation functions (FFs) to take into account medium-induced energy loss are
able to reproduce the suppression in central collisions with $dE/dx\sim$ 0.25 GeV/fm
(equivalent to $dE/dx\sim$ 7 GeV/fm in a static medium, much larger than the standard
energy losses found in {\it cold} nuclear matter \cite{enterria:francois}).
Peripheral collision spectra, on the other side, are consistent with $dE/dx\sim$ 0. 
PHENIX central $AuAu$ high-$p_T$ results are thus in qualitative agreement with the
expectations of parton energy loss in an opaque medium.

\section{High-$p_T$ hadron composition.}

The spectra \cite{enterria:julia} of identified $\pi^\pm$ and $p,\bar{p}$ in central $Au+Au$ collisions
cross each other at about $p_T =$ 2 GeV/c and give a meson to baryon yield ratio of $\sim$1. 
Such an observation is at variance with $NN$ data (ISR, FNAL) where a baryon/meson ratio of 
$\sim$0.3 - 0.4 was found at high-$p_T$. Various interpretations \cite{enterria:julia} have been 
proposed to explain such a behaviour ranging from large strong collective radial flow 
(proportional to the hadron mass) plus hadron rescattering, to more ``exotic'' mechanisms 
such as baryon junctions and jet quenching in the pion channel.


\section{Electron yield and charm production.}

\begin{figure}
\psfig{figure=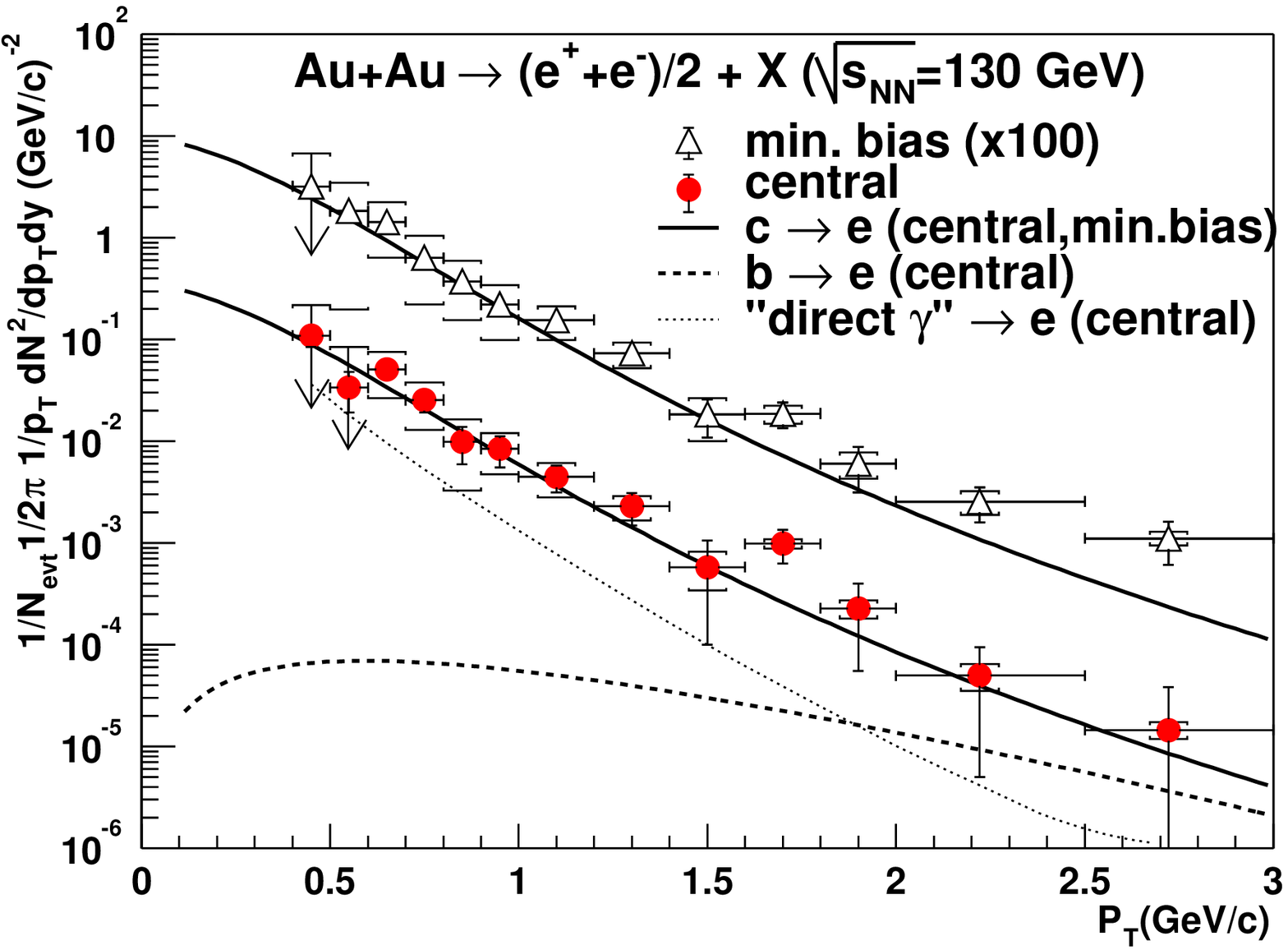,height=2.35in}
\psfig{figure=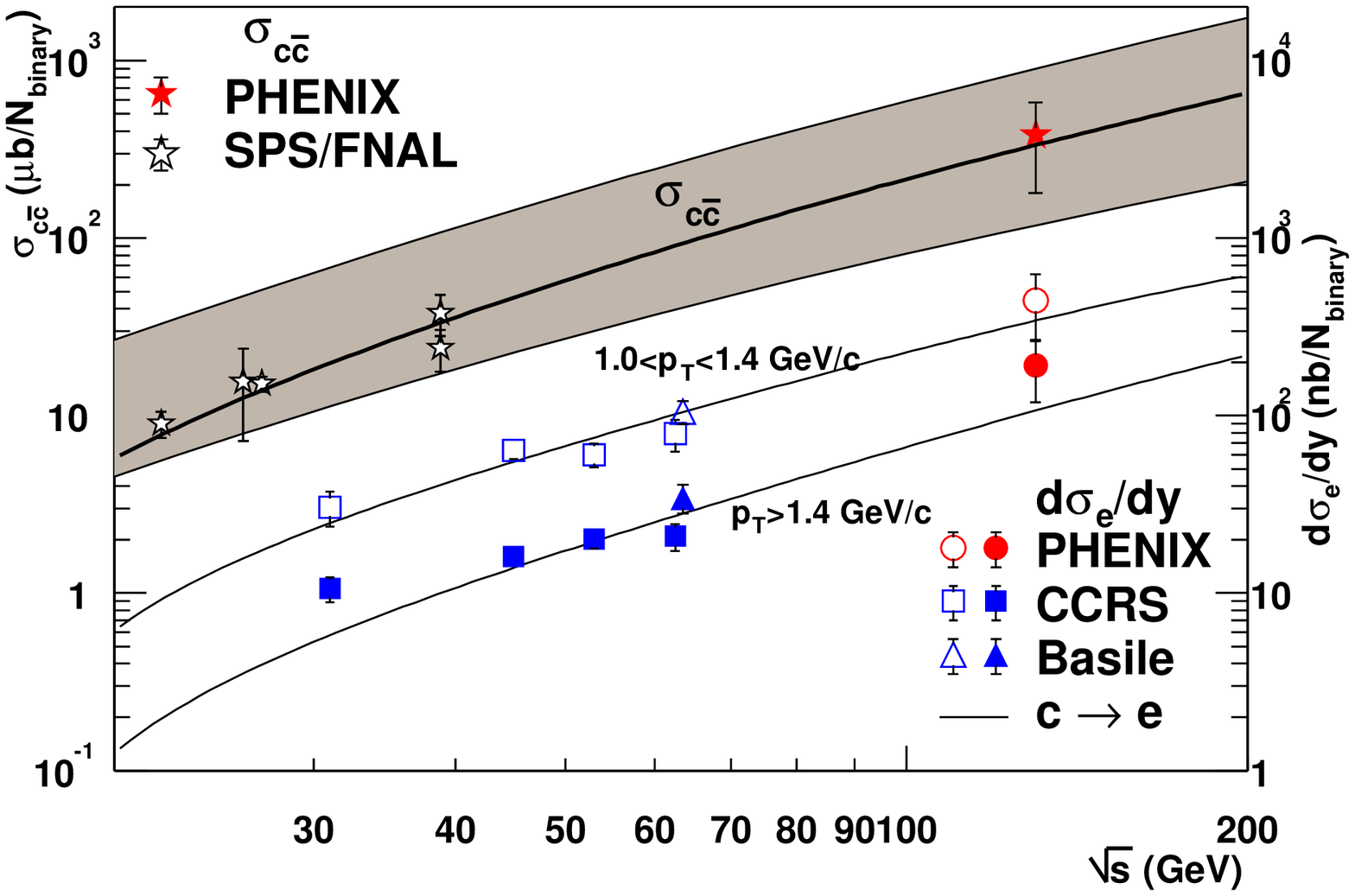,height=2.35in}
\caption{Left: Inclusive electron distribution for central and min. bias $Au+Au$ collisions,
after subtraction of the light meson decay background, compared to the scaled $pp$ contribution 
from open charm meson decays. Right: Inclusive charm cross-section as a function
of $\sqrt{s}$ from experimental data, pQCD and PYTHIA calculations \protect\cite{enterria:akiba}.
\label{fig:3}}
\end{figure}

Single electrons are very interesting observables providing information on
several hard probe signals: heavy-flavor (via semi-leptonic decays of open charm and bottom
mesons), Drell-Yan/thermal dileptons, and direct $\gamma$ (through conversion).
The inclusive electron spectrum, however, is dominated by light meson backgrounds:
Dalitz decays of pseudo-scalars ($\pi^0$ and $\eta$), soft $\gamma$ conversions, 
and di-electron decays of vector mesons. Fig. \ref{fig:3} shows the single electron, $(e^-+e^+)/2$,
spectra measured by PHENIX \cite{enterria:akiba} in the $p_T$ range 0.2 - 3.0 GeV/c after 
subtraction of the ``cocktail'' background. The resulting spectra for
minimum bias and central collisions are consistent with scaled $pp$ Pythia calculations
of open charm contributions. Assuming no thermal radiation component (virtual and/or 
real+conversion) in the spectra, i.e. assuming all $e^\pm$ originate in $D$ decays,
the obtained charm cross-section (per $NN$ collision) is $\sigma_{c\bar{c}}=380 \pm 60 \pm 200$ $\mu$b
for the central sample in agreement with the expectation from binary scaling of $pp$ data and
with theoretical calculations (Fig. \ref{fig:3} right). Thus, assuming a modest
photon contribution, the measured electron spectra and cross-sections are consistent with simple 
$pp$ scaled charm production precluding significant nuclear or medium effects on heavy-flavor production.

{\bf Acknowledgments}: I am thankful to the members of {\it Groupe Photons}
at Subatech (G.~Mart\'{\i}-nez, H. Delagrange, Y. Schutz and L. Aphecetche) for fruitful 
discussions. Support of the 5th European Union TMR Programme through Marie-Curie 
Fellowship No. HPMF-CT-1999-00311 is acknowledged.

\end{document}